\documentclass[12pt,preprint]{aastex}

\usepackage{color}

\begin{document}
\title{Lagrangian Statistics of Dark Halos in a $\Lambda$CDM Cosmology}
\author{\sc Jounghun Lee\altaffilmark{1}, Oliver Hahn\altaffilmark{2}
and Cristiano Porciani\altaffilmark{2,3}}
\altaffiltext{1}{Department of Physics and Astronomy, FPRD, 
Seoul National University, Seoul 151-747 , Korea}
\altaffiltext{2}{Institute for Astronomy, ETH Zurich, CH-8093 Z\"urich, 
Switzerland}
\altaffiltext{3}{Argelander Institut f\"ur Astronomie, Auf dem H\"ugel 71, 
D-53121 Bonn, Germany}
\email{jounghun@astro.snu.ac.kr}
\begin{abstract}
New statistical properties of dark matter halos in Lagrangian space are 
presented. Tracing back the dark matter particles constituting bound halos
resolved in a series of N-body simulations,  
we measure quantitatively the correlations of the proto-halo's inertia 
tensors with the local tidal tensors and investigate how the correlation 
strength depends on the proto-halo's sphericity, local density and filtering 
scale.  It is shown that the majority of the proto-halos exhibit strong 
correlations between the two tensors provided that the tidal field is smoothed 
on the proto-halo's mass scale. The correlation strength is found to increase 
as the proto-halo's sphericity increases, as the proto-halo's mass increases, 
and as the local density becomes close to the critical value, $\delta_{ec}$.
It is also found that those peculiar proto-halos which exhibit exceptionally  
weak correlations between the two tensors tend to acquire higher specific 
angular momentum in Eulerian space, which is consistent with the linear tidal 
torque theory. In the light of our results, it is intriguing to speculate a 
hypothesis that the low surface brightness galaxies observed at present 
epoch correspond to the peculiar proto-halos with extreme low-sphericity 
whose inertia tensors are weakly correlated with the local tidal tensors.
\end{abstract}

\section{INTRODUCTION}

Galaxies are biased tracers of the underlying dark matter distribution.
It has been well known that the two-point correlation function of the observed 
galaxies follows a power-law, different from that of the dark matter 
determined in N-body simulations \citep[e.g.,][]{mad-etal90}. 
To test theoretical predictions based on the dark matter against real 
observations of galaxies, it is required to determine a hidden connection 
between the galaxies and dark matter. The density peak formalism was the 
first attempt to provide such a connection, according to which the galaxies 
form in the high-peaks (i.e., local maxima) of the initial smoothed density 
field \citep{dav-etal85,bar-etal86,kai86,bon-mye96}.  Fitting quite 
well into the cold dark matter paradigm and explaining successfully some 
observed properties of the galaxies, this density peak formalism became the 
most prevalent model for biased galaxy formation. Given that 
gravity is mainly responsible for the formation and evolution of 
the galaxies, it was indeed reasonable, appropriate, and natural to 
regard the initial density peaks as good indicators of the sites of galaxy 
formation.

An uncomfortable truth, however, had to be faced when \citet{kat-etal93} 
reported the unexpected results derived from N-body simulations that 
the dark matter particles of galactic halos are not well overlapped with 
those from the high-peaks of the initial density field. 
They traced the trajectories of dark matter particles that comprise 
the initial density peaks and found that the particles from 
the initial peaks do not end up in real halos. Furthermore, it was also shown 
that the correlation function of high density peaks is 
different from that of galactic halos. These disturbing results forced 
them to conclude that the high-peaks of the initial density field cannot 
be good indicators of the sites of galaxy formation. 
In the same spirit, \citet{por-etal02b} have shown that only $\sim 40\%$ of 
the proto-halos in a N-body simulation contain a density peak within their 
Lagrangian volume and that gravitational shear plays an important role in 
shaping the proto-halos.

To find a good indicator of the initial sites of galaxy formation, it is first 
necessary to understand the statistical properties that the majority of the 
proto-galactic sites possess in Lagrangian space. Such a property was first 
found unwittingly by \citet{lee-pen00} while studying the origin of the 
galaxy angular momentum. 
According to the linear tidal torque theory \citep{dor70,whi84}, 
the angular momentum of a proto-galaxy is generated at first order only   
when the local tidal shear tensor is not perfectly correlated with the 
inertia momentum tensor of the proto-galaxy. In previous works dealing 
with the linear tidal torque theory \citep[e.g.,][]{cat-the96}, it was 
assumed that the two tensors are generally uncorrelated. \citet{lee-pen00} 
tested the validity of this assumption against the numerical data from N-body 
simulations. They calculated the correlations between the two 
tensors at proto-halo sites and found for the first time that the two 
tensors are in fact quite strongly correlated with each other in contrast to 
the general assumption. Their result, however, was based on rather 
low-resolution N-body simulations and thus failed to draw serious attentions. 

Later, \citet{por-etal02b} re-tested this assumption against high-resolution 
simulations, confirming and extending the preliminary results of 
\citet{lee-pen00}. Noting that the existence of strong correlations between 
the two tensors is such a common phenomenon exhibited by most of the 
proto-halos they considered, \citet{por-etal02b} have suggested that 
the boundaries of the proto-halos are determined by the push and pull
of the external mass distribution. This is in some sense the opposite of
the common wisdom where the key factor is assumed to be the self-gravity
attraction. 

In this paper we want to further improve upon the previous work by 
\citet{lee-pen00} and \citet{por-etal02b} by using a more detailed statistical 
treatment and simulations of better quality. Our goal here is to quantify the 
correlations between the two tensors as new Lagrangian statistics of 
dark halos and investigate how the correlation strengths depend on 
proto-halo's shapes and local environments. The outline of this paper is as 
follows. In \S 2, a brief description of N-body data is provided and the 
statistical analysis of it is presented. In \S 3 is presented new Lagrangian 
statistics of dark halos related to the correlations of the proto-halo's 
inertia tensors with the local tidal tensors. In \S 4 an implication of our 
result on the low surface brightness galaxies is explained. In \S 5 the 
results are discussed and a final conclusion is drawn. Throughout this paper, 
we assume a spatially flat $\Lambda$CDM cosmology.
 
\section{DATA AND ANALYSIS}

We use the samples of dark matter halos obtained from three different N-body 
simulations conducted by \citet{hah-etal07} in periodic boxes of linear size 
$L_{1}=45h^{-1}$Mpc, $L_{2}=90h^{-1}$Mpc and $L_{3}=180h^{-1}$Mpc. 
All three simulations assume a spatially flat $\Lambda$CDM cosmology with 
$\Omega_{m}=0.25$, $\Omega_{\Lambda}=0.75$, $\Omega_{b}=0.045$, 
$\sigma_{8}=0.9$, $H_{0}=0.73$ and $n_{s}=1$, each following the evolution 
of $512^{3}$ particles to the present epoch from a given initial redshift 
$z_{i}$ ($L_{1}$, $L_{2}$ and $L_{3}$ have $z_{i}=79$, $z_{i}=65$ 
and $z_{i}=52$, respectively). Bound halos were resolved in each simulation 
by applying the standard friends-of-friends algorithm with linking-length 
parameter of $b=0.2l$ here $l$ denotes the mean inter-particle distance 
\citep{efs-etal85}. Among them, only those halos comprising more than $300$ 
dark matter particles were selected to avoid possible numerical artifacts. 
A total of $50839$ halos are selected ($13390$, $16339$ and $21110$ halos 
from the $L_{1}$, $L_{2}$ and $L_{3}$ simulations, respectively), which 
span a wide mass range of $[10.2, 15.3]$ in units of $h^{-1}M_{\odot}$. 
A full description of the simulations and the process of the 
halo-identification is provided in \citet{hah-etal07}.

Tracing back to the initial conditions the trajectories of dark matter 
particles that constitute each selected halo, we locate the proto-halo sites 
in the Lagrangian space corresponding to the initial epoch $z_{i}$. Then, we 
determine the center of mass of each proto-halo site using the positions of 
its constituent particles. At each halo's center of mass, we measure the mean 
density contrast $\delta$ in the initial density field smoothed with a top-hat 
filter of scale radius $R_{s}$. Here we consider four different filtering 
scales: $R_{s}=0.5$, $1$, $2$ and $5\, h^{-1}$Mpc.
The initial peculiar potential field $\phi$ was derived from the density field 
by solving the Poisson equation \citep{hah-etal07}, and the tidal shear 
tensor ${\bf T}\equiv (T_{ij})$ was measured as the second derivative of 
$\phi$ at the center of mass of each proto-halo. Diagonalizing ${\bf T}$, 
we determine the three eigenvectors $\{{\bf t}_{1},{\bf t}_{2},{\bf t}_{3}\}$ 
corresponding to the three eigenvalues 
$\{\lambda_{1},\lambda_{2},\lambda_{3}\}$ in a decreasing order. 

The inertia momentum tensor ${\bf I}\equiv (I_{ij})$ of each proto-halo site 
is also determined using the positions of the constituent particles in 
accordance with the formula given in \citet{hah-etal07}:
\begin{equation}
\label{eqn:ine}
I_{ij} \equiv m_{\alpha}\sum_{\alpha}\left(r_{\alpha}^2
\delta_{ij}-x_{\alpha,j}x_{\alpha,k}\right),
\end{equation}
where $m_{\alpha}$ is the mass of $\alpha$-th particle, ${\bf x}_{\alpha}$ 
is the position vector of the $\alpha$-th particle from the center of mass 
of the proto-halo site, and $\delta_{jk}$ is the Kronecker symbol. 
When the correlations with the tidal field smoothed on the scale $R_{s}$ 
are calculated, we consider only those halos whose mass $M$ is in the range 
of $0.9M_{s}< M <1.1M_{s}$ where  $M_{s}$ is the top-hat mass enclosed by 
the radius $R_{s}$, given the fact that a given proto-halo with mass $M$ 
would experience the strongest effect from the tidal field smoothed 
on the same mass scale $M=M_{s}$ \citep[][see also the appendix in 
\citet{hah-etal09}]{lee-pen00}. Table \ref{tab:list} 
lists the number ($N_{h}$) and mean mass ($\bar{M}$) of halos considered for 
the correlations with the tidal field smoothed on the four different 
filtering radii $R_{s}$.
 
A similarity transformation is performed to reexpress ${\bf I}$ in the 
principal frame of ${\bf T}$. To quantify the degree of the correlations 
between the two tensors, we define a parameter $\beta$, as
\begin{equation}
\label{eqn:beta}
\beta \equiv 1 -
\left(\frac{\varrho^{2}_{12}+\varrho^{2}_{23}+\varrho^{2}_{31}}
{\varrho^{2}_{11}+\varrho^{2}_{22}+\varrho^{2}_{33}}\right)^{1/2},
\end{equation}
where $\{\varrho_{11},\varrho_{22},\varrho_{33}\}$ and 
$\{\varrho_{12},\varrho_{23},\varrho_{31}\}$ represent the three diagonal 
and off-diagonal elements of ${\bf I}$ in the principal frame of 
${\bf T}$, respectively. If a proto-halo region has a perfectly 
spherical shape, then the eigenvectors of its inertia tensor are degenerate 
(i.e., any axis frame can be its eigenvector system) and all of the 
off-diagonal elements of ${\bf I}$ are always zero 
($\varrho_{12}=\varrho_{23}=\varrho_{31}=0)$ .Thus for the case of 
a perfectly spherical proto-halo region, we always have $\beta=1$.
When a proto-halo region is not perfectly spherical but its inertia tensor 
${\bf I}$ is perfectly correlated with the tidal shear tensor ${\bf T}$ 
measured at its center of mass, then ${\bf I}$ should be completely diagonal 
in the principal axis frame of ${\bf T}$. Thus for the case of a perfect 
correlation between ${\bf I}$ and ${\bf T}$, we will also have $\beta =1$. 
On the other hand, if the two tensors are uncorrelated, then ${\bf I}$ in 
the principal axis frame of ${\bf T}$ is not diagonal and the off-diagonal 
elements should be as large as the diagonal ones in magnitude unless the 
eigenvectors of ${\bf I}$ are degenerate. As the strength of the correlations 
between ${\bf I}$ and ${\bf T}$ decreases,  the degree of the deviation of 
$\beta$ from the value of unity will increase.

In the following section, we determine the probability distribution of 
$\beta$ and investigate how the value of $\beta$ depends on the proto-halo's 
shape, local density and filtering radius.

\section{CORRELATIONS BETWEEN INERTIA AND TIDAL TENSORS}

\begin{figure}
  \begin{center}
   \plotone{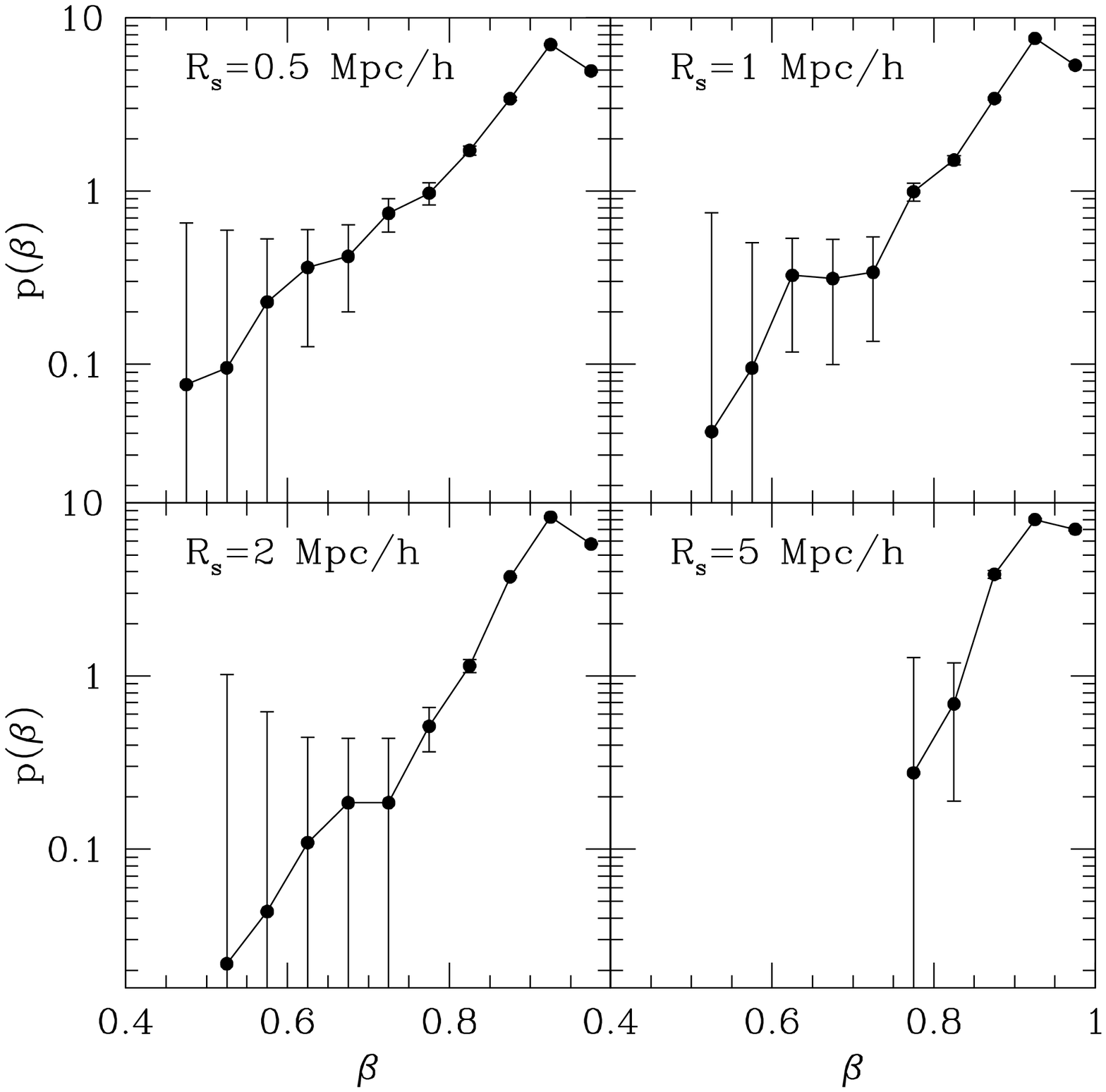}
\caption{Probability density distribution of $\beta$ of the proto-galactic 
sites with Poissonian errors when the initial density field is smoothed on 
the scale of $0.5,\ 1,\ 2$ and $5\, h^{-1}$Mpc (top-left, top-right, 
bottom-left, and bottom-right, respectively).}
\label{fig:pro}
 \end{center}
\end{figure}

Using the numerical data described in \S 2, we first determine the probability 
density distribution, $p(\beta)$. Figure \ref{fig:pro} shows $p(\beta)$ for 
the four different cases of $R_{s}=0.5$, $1$, $2$ and $5\, h^{-1}$Mpc 
(top-left, top-right, bottom-left, and bottom-right  panel, respectively). 
It is worth mentioning again here that when we calculate $\beta$ based on the 
tidal field smoothed on the scale $R_{s}$ we consider only those proto-halos 
whose masses belong to the range $(0.9M_{s},1.1M_{s})$ with 
$M_{s}\equiv (4\pi/3)\bar{\rho}R^{3}_{s}$, where $\bar{\rho}$ is the mean 
mass density of the universe. In each panel the errorbars represent 
the Poissonian noise. In all cases, the distribution $p(\beta)$ is strongly 
biased towards the high-$\beta$ section, reaching a maximum at $\beta\ge 0.9$. 
It is now clear that at the proto-halo sites the tidal shear and inertia 
momentum tensors are strongly correlated with each other, regardless of the 
smoothing scale, which confirms quantitatively the previous works 
\citep{lee-pen00,por-etal02b}. Note also that for the case of 
$R_{s}=0.5\, h^{-1}$Mpc a small number of proto-halos exhibit exceptionally 
low values of $\beta \le 0.5$, while for the cases of 
$R_{s}=1,\ 2,\ 5\, h^{-1}$Mpc all proto-halos have $\beta \ge 0.5$. 
 
This result also implies that a halo of mass $M$ tends to be made of those 
dark matter particles that accreted into the initial sites along the 
principal axes of the local tidal field smoothed on the same mass scale $M$, 
which is consistent with the Zel'dovich approximation \citep{zel70}.

\subsection{Dependence on Protohalo's Sphericity and Linear Density}

\begin{figure}
  \begin{center}
   \plotone{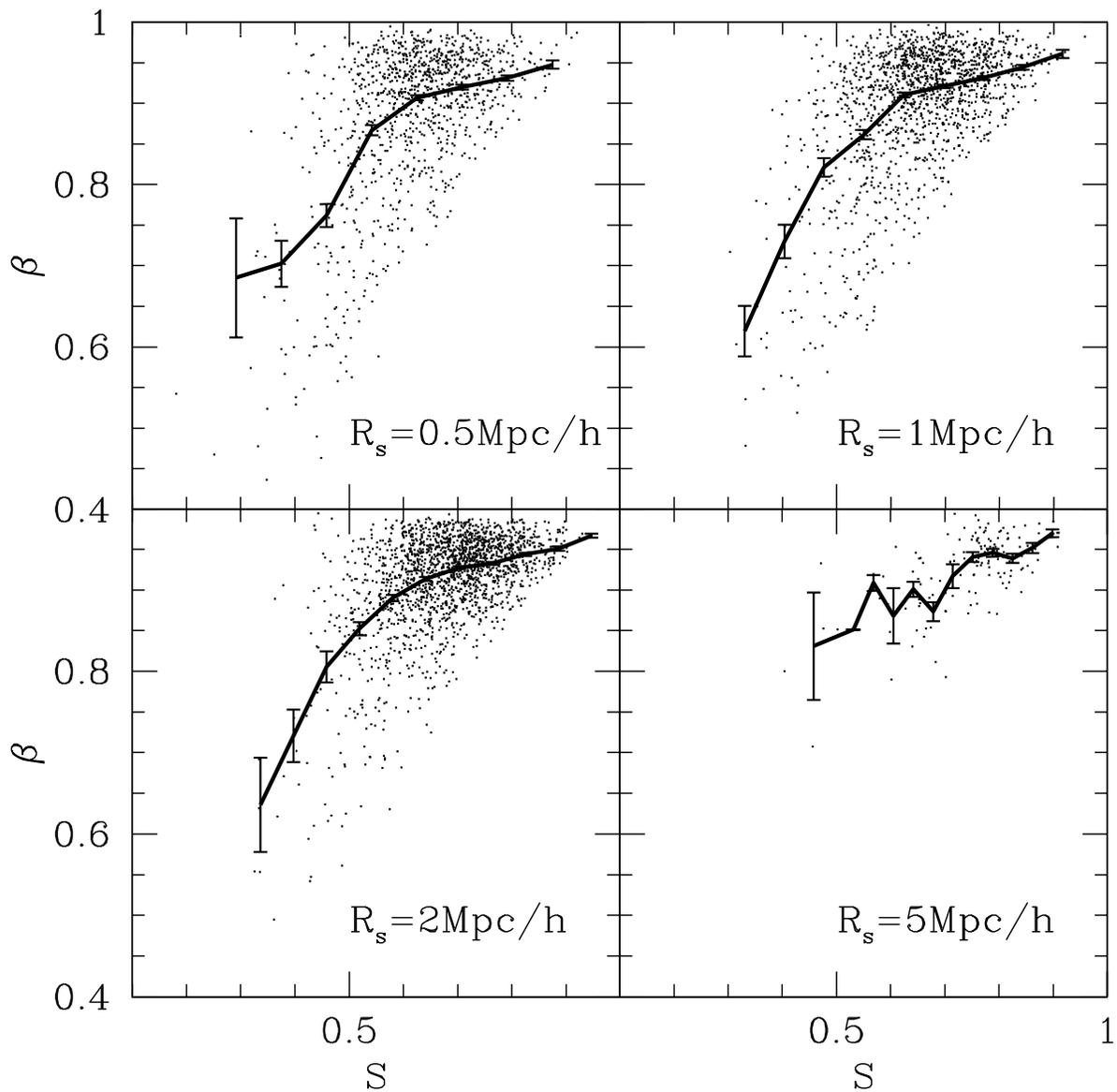}
\caption{Scatter plots of $\beta$ vs. the sphericity $S$ of the 
proto-halo sites when the initial density field is smoothed on 
the scale of $0.5,\ 1,\ 2$ and $5\, h^{-1}$Mpc (top-left, top-right, 
bottom-left, and bottom-right, respectively). In each panel, the solid 
line represents the mean values $\langle\beta\rangle$ and the errorbars 
indicate the standard deviation in the measurement of $\langle\beta\rangle$.}
\label{fig:shape}
 \end{center}
\end{figure}

We investigate how $\beta$ varies with the shapes of the proto-halos.
Using the three eigenvalues of ${\bf I}$, we measure the sphericity $S$ of 
each proto-halo as \citep{hah-etal07}
\begin{equation}
S\equiv\left(\frac{\varrho_3}{\varrho_1}\right)^{1/2},
\end{equation}
where $\varrho_1$ and $\varrho_3$ represent the largest and the smallest 
eigenvalue of ${\bf I}$, respectively. Binning the range of $S$, we measure  
the mean of $\beta$ averaged over a given differential bin, $[S,S+dS]$. 
Figure \ref{fig:shape} shows the scatter-plot of $\beta$ vs. $S$. The thick 
solid line corresponds to the mean value $\langle\beta\rangle$ as a function 
of $S$. The errors represent one standard deviation in the measurement of 
$\langle\beta\rangle$ calculated as 
$\langle\Delta\beta^{2}\rangle/\sqrt{(n-1)}$ where $n$ represents the number 
of the proto-halos belonging to a given differential bin $[S,S+dS]$.
It can be noted that $\beta$ increases almost monotonically with $S$.
In other words, the less spherical a proto-halo is, the weaker is the
correlation between the tidal shear and the inertia momentum tensors. 
For those proto-halos with low sphericity ($S\le 0.4$), 
the mean value, $\langle\beta\rangle$, drops below $0.7$. 

\begin{figure}
  \begin{center}
   \plotone{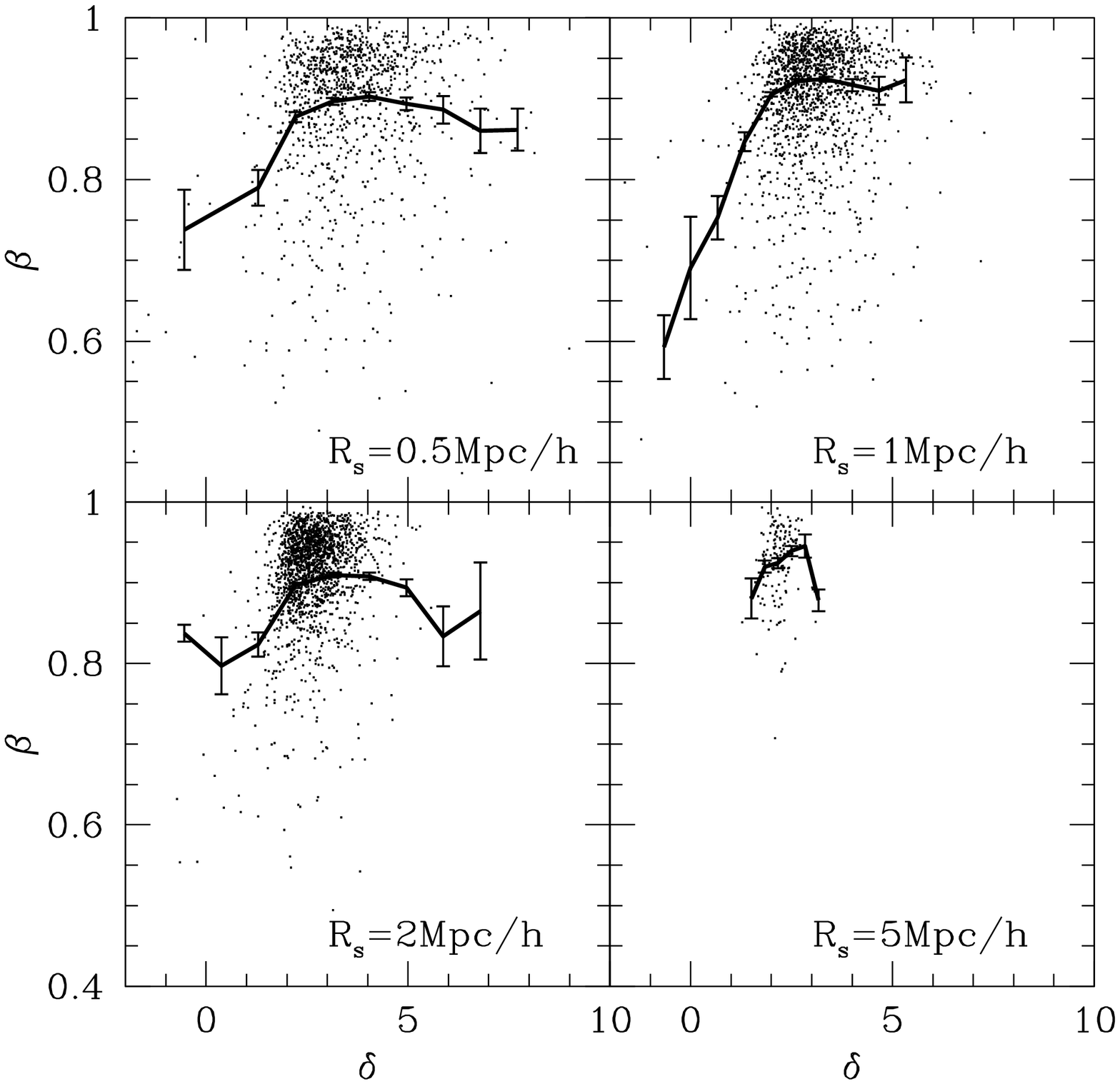}
\caption{Scatter plots of $\beta$ vs. the linearly extrapolated density 
$\delta$ of the proto-halo sites when the initial density field is smoothed 
on the scale of $0.5,\ 1,\ 2$ and $5\, h^{-1}$Mpc (top-left, top-right, 
bottom-left, and bottom-right, respectively). In each panel, the solid 
line represents the mean values $\langle\beta\rangle$ and the 
errors correspond to the one standard deviation in the measurement of 
$\langle\beta\rangle$.}
\label{fig:den}
 \end{center}
\end{figure}

We also investigate how $\beta$ changes with the local density field at the 
proto-halo sites. Let $\delta_{i}$ denote the initial density contrast 
measured at the proto-halo's center of mass. Since the three simulations 
used here started at different initial redshifts $z_{i}$ (see \S 2), we 
use the linearly extrapolated linear density $\delta$ to $z=0$ instead of 
$\delta_{i}$ itself, which is calculated as 
$\delta_{L}\equiv [D(z)/D(0)]\delta_{i}$ where $D(z)$ is the linear growth 
factor.  We bin the range of $\delta$ and calculate the mean of $\beta$ 
averaged over a given differential bin, $[\delta, \delta+d\delta]$.
Figure \ref{fig:den} shows the scatter-plots of $\beta$ vs. $\delta$. 
The thick solid line corresponds to the mean value $\langle\beta\rangle$ 
as a function of $\delta$. Note first that for the case of the large filtering 
radius $R_{s}=5h^{-1}$Mpc all values of $\delta$ lie in quite a narrow range 
converging to the critical density value $\delta_{ec}\approx 2.5$ for the 
ellipsoidal collapse \citep{she-etal01,des08,rob-etal09}. In contrast, for 
the cases of smaller filtering radii $R_{s}=0.5,\ 1,\ 2\,h^{-1}$Mpc, the 
values of $\delta$ spread over quite a wide range from $-1$ to $10$. 
The value of $\beta$ tends to be highest when $\delta$ becomes close to 
$\delta_{ec}$. As $\delta$ deviates from $\delta_{ec}$, the value of 
$\beta$ decreases both in the high-$\delta$ and in the low-$\delta$ 
sections.

\begin{figure}
  \begin{center}
   \plotone{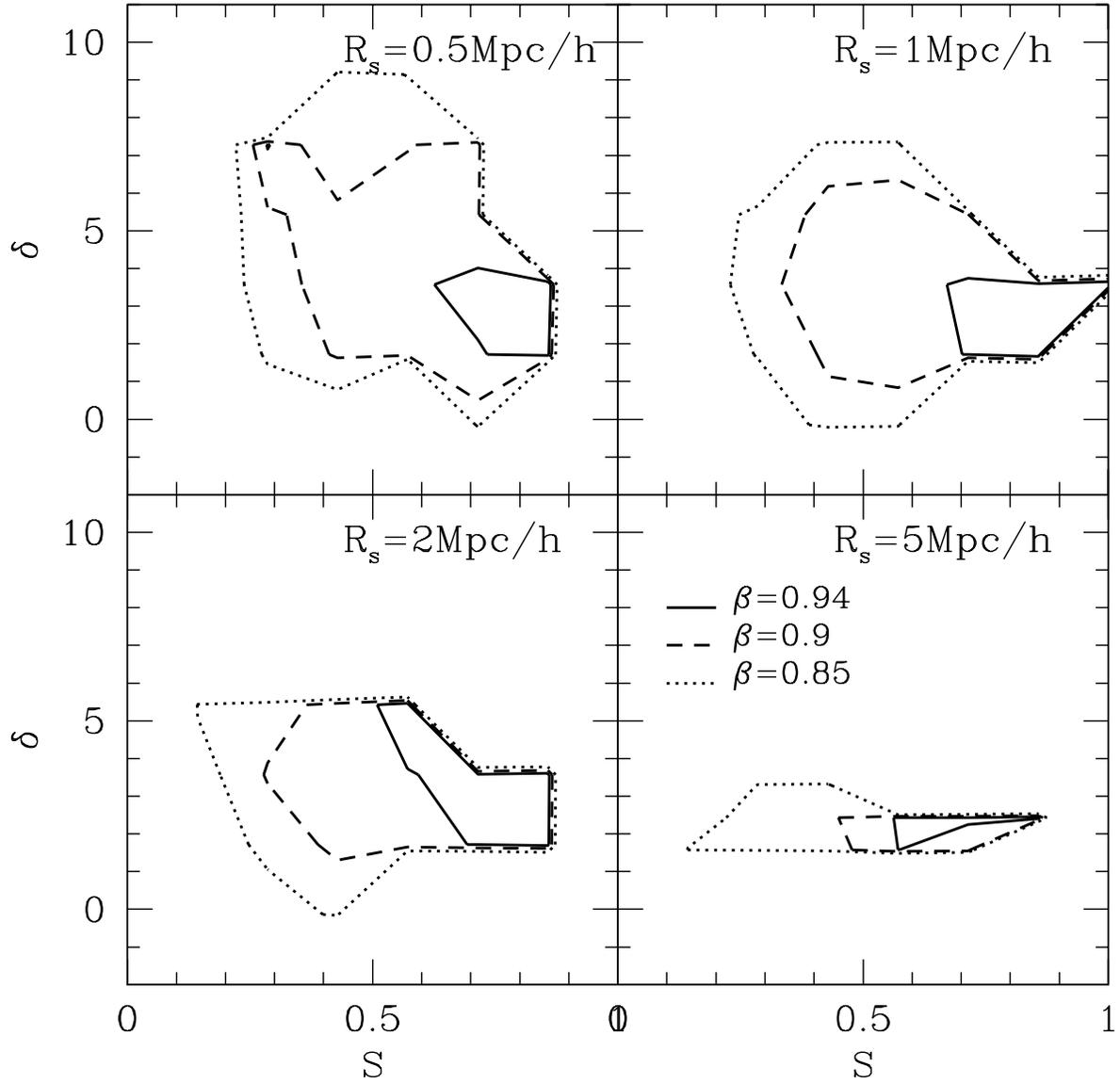}
\caption{Contour plots of $\beta$ in the $S$-$\delta$ when the initial 
density field is smoothed on the scale of $0.5,\ 1,\ 2$ and $5\, h^{-1}$Mpc 
(top-left, top-right, bottom-left, and bottom-right, respectively).}
\label{fig:con}
 \end{center}
\end{figure}

Since $S$ and $\delta$ are not mutually independent but $S$ tends decrease 
in the low-$\delta$ region, it may be interesting to see how $\beta$ vary 
as $S$ and $\delta$ change simultaneously. 
Figure \ref{fig:con} shows the contour-plots of $\beta$ in the 
$S$-$\delta$ plane. This plot clearly demonstrates that those proto-halos 
with high $S$ and $\delta\sim \delta_{ec}$ tend to have a high value of 
$\beta$.

\subsection{Dependence on Filtering Scale}

We now explore the dependence of $\beta$ on the filtering radius. 
Let us consider two different scales $R_{\rm th}$ and $R_{s}$. We first 
calculate the inertia tensors of those halos whose masses lie in a fixed 
range of $(0.9M_{\rm th},1.1M_{\rm th})$ where $M_{\rm th}$ denotes the 
top-hat masses enclosed by the filtering scale $R_{\rm th}$. Then, we 
calculate the mean correlations $\langle\beta\rangle$ of these 
inertia tensors with the tidal tensors smoothed on the different scale 
of $R_{s}$. For each given $R_{\rm th}$, we consider four different scales 
$R_{s}$ for the smoothing of the tidal tensors. 

\begin{figure}
  \begin{center}
   \plotone{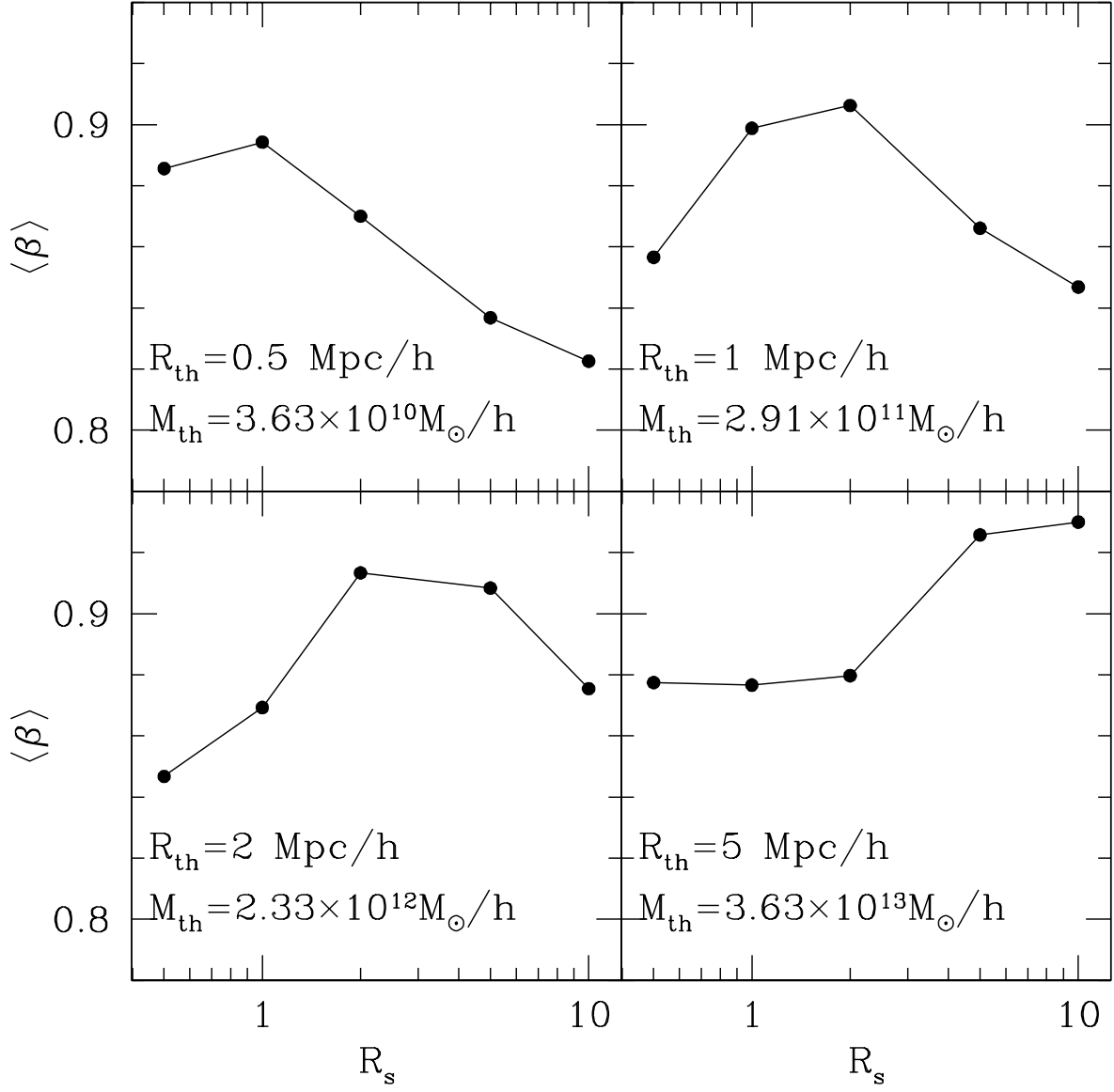}
\caption{Mean values $\langle\beta\rangle$ as a function of the filtering 
radius $R_{s}$ for four different values of the proto-halo masses. The 
four masses are enclosed by the four top-hat radii $R_{\rm th}$ as 
$M_{\rm th}=(4\pi/3)\bar{\rho}R^{3}_{\rm th}$ where 
$R_{\rm th}=0.5,\ 1,\ 2$ and $5\, h^{-1}$Mpc (top-left, top-right, 
bottom-left, and bottom-right, respectively).}
\label{fig:fil}
 \end{center}
\end{figure}

Figure \ref{fig:fil} shows $\langle\beta\rangle$ as a 
function of the filtering scale $R_{s}$. Each panel plots the mean values of 
$\beta$ calculated using the inertia tensors of the proto-halos with mass 
$M_{\rm th}$ enclosed by a fixed  radius $R_{\rm th}$ and the tidal tensors 
smoothed on four different scales $R_{s}$. 
It can be seen that $\langle\beta\rangle$ increases monotonically when 
$R_{s}\le 2R_{\rm th}$ and decreases sharply when $R_{s}> 2R_{\rm th}$.
The value of $\beta$ reaches a maximum when 
$R_{\rm th}\le R_{s}\le 2R_{\rm th}$. In terms of mass, it can be 
said that $\beta$ becomes maximal at $M_{\rm th}\le M_{s}\le 8M_{\rm th}$.
In other words, the inertia tensors of the proto-halos with masses 
$M_{\rm th}$ are strongly correlated with the local tidal fields 
smoothed on the mass scales that have the same order of magnitude as 
$M_{\rm th}$.  If the tidal fields are smoothed on the scales order of 
magnitude smaller or larger than $M_{\rm th}$, the correlations between 
the two tensors decrease. Thus, the correlation strength between 
the two tensors depends on the scale on which the tidal field is smoothed. 

\section{IMPLICATION ON THE LOW SURFACE BRIGHTNESS GALAXIES}

\begin{figure}
  \begin{center}
   \plotone{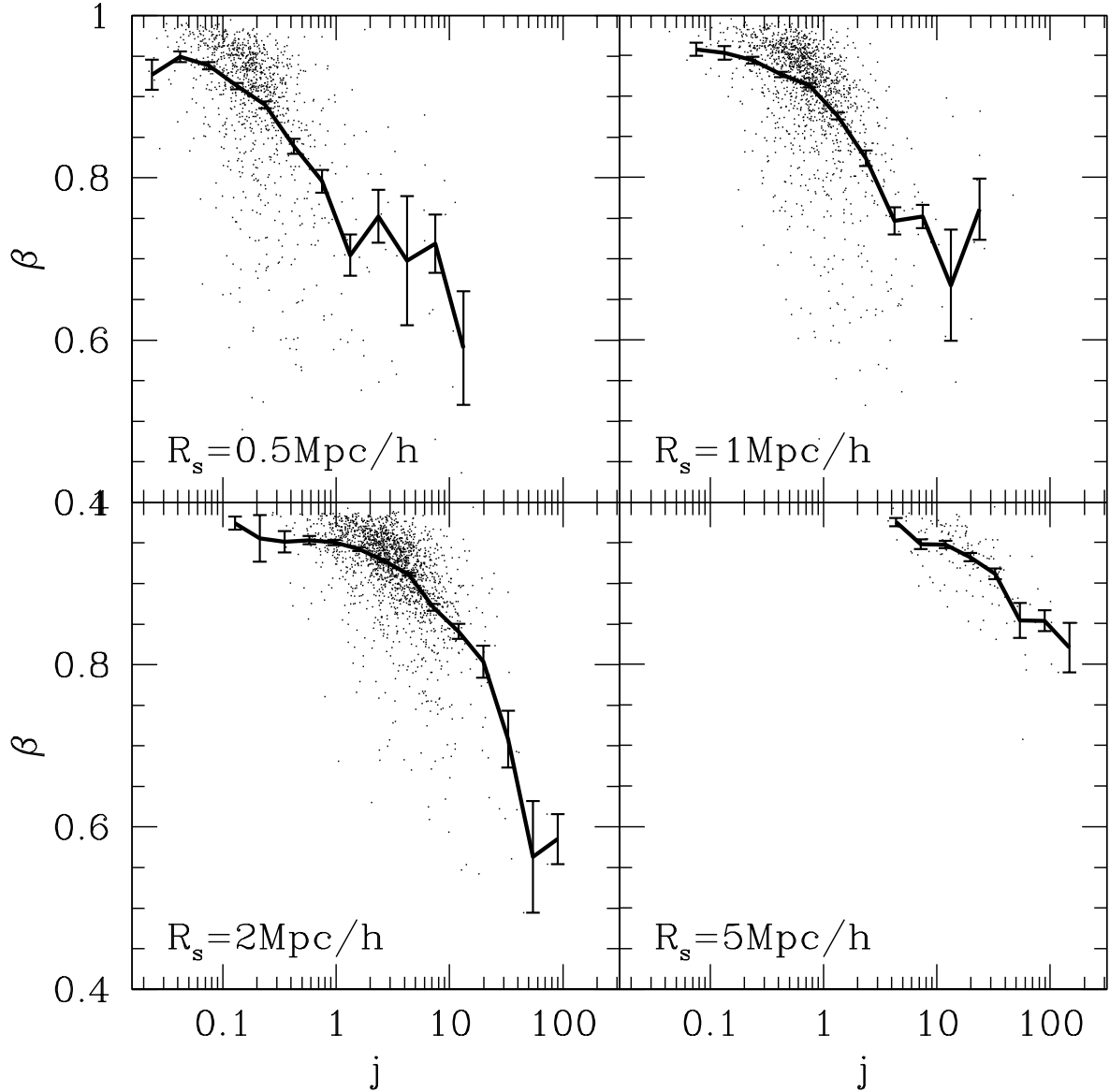}
\caption{Scatter plots of $\beta$ vs. the specific angular momentum 
$j$ of the proto-halo sites when the initial density field is smoothed 
on the scale of $0.5,\ 1,\ 2$ and $5\, h^{-1}$Mpc (top-left, top-right, 
bottom-left, and bottom-right, respectively). In each panel, the solid 
line represents the mean values $\langle\beta\rangle (j)$ and the 
errors correspond to the one standard deviation in the measurement of 
$\langle\beta\rangle (j)$.}
\label{fig:ang}
 \end{center}
\end{figure}

In the light of our results, it is time to recall the linear tidal torque 
theory according to which the magnitude of the specific angular momentum 
(angular momentum per unit mass) of a proto-halo increases as the correlation 
of its inertia tensor with the local tidal tensor increases \citep{dor70,whi84,
cat-the96,lee-pen00}. In the context of the linear tidal torque theory, those 
proto-halo sites which have low values of $\beta$ are likely to acquire higher 
specific angular momentum \citep{por-etal02a}. To test this core prediction 
of the linear tidal torque theory, we explore the dependence of $\beta$ on 
the specific angular momentum measured in Eulerian space. For each halo, we 
first measure the angular momentum vector ${\bf J}$ in the Eulerian space as  
\begin{equation}
\label{eqn:ang}
{\bf J} \equiv \sum_{\alpha}m_{\alpha}{\bf r}_{\alpha}\times {\bf v}_{\alpha},
\end{equation}
where ${\bf r}_{\alpha}$ and ${\bf v}_{\alpha}$ represent the position and 
the velocity of the $\alpha$-th particle in the halo's center of mass frame, 
respectively. The specific angular momentum vector ${\bf j}$ of each halo 
is then calculated as ${\bf j}\equiv {\bf J}/M$.
Binning the range of $j$, we calculate the mean of $\beta$ averaged over a 
given differential bin, $[j,j+dj]$. Figure \ref{fig:ang} shows the 
scatter-plot of $\beta$ vs. $j$. The thick solid line corresponds to the mean 
value $\langle\beta\rangle (j)$ and the errors represent one standard 
deviation in the measurement of $\langle\beta\rangle (j)$. As can be seen, 
there is a strong trend that $\beta$ increases with $j$. In other words, 
those proto-halos whose inertia tensors are less correlated with the local 
tidal tensors are likely to acquire higher specific angular momentum 
vectors in the subsequent evolution. 

A crucial implication of our results is that the proto-halos with the lowest 
$S$ will thus acquire the highest specific angular momentum since the 
proto-halos with the lowest $S$  are found to have the weakest correlations 
between the inertia and tidal shear tensors (see Figure \ref{fig:shape}).  
It is very interesting to recall that the high-specific angular momentum is 
the characteristic property of the low-surface brightness galaxies (LSBGs) 
\citep[][and references therein]{boi-etal03,mon-etal03}. Our result leads 
to a speculation that the LSBGs might originate from those peculiar 
proto-halos with very low values of $S$ whose inertia tensors are weakly 
correlated with the local tidal tensors. 

It is, however, worth mentioning here that the low values of $S$ do not 
necessarily correspond to elongated particle distribution of proto-halo 
regions. A proto-halo region can have low values of $S$ if it consists of 
two disconnected patches with center of mass located in the middle.
To examine whether or not the proto-halos with low values of $S$ have 
connected particle distribution, we inspect the particle distribution of 
those proto-halos with $S$ below the tenth percentile in the mass range 
$2\le M/[10^{11}\, h^{-1}M_{\odot}]\le 4$. We consider only the low-mass halos 
for this inspection since they exhibit the lowest values of $S$. 
It is found that among the inspected proto-halos, approximately two third of 
them have connected particle distribution while the other one third consist 
of more than two patches. 

\begin{figure}
  \begin{center}
   \plotone{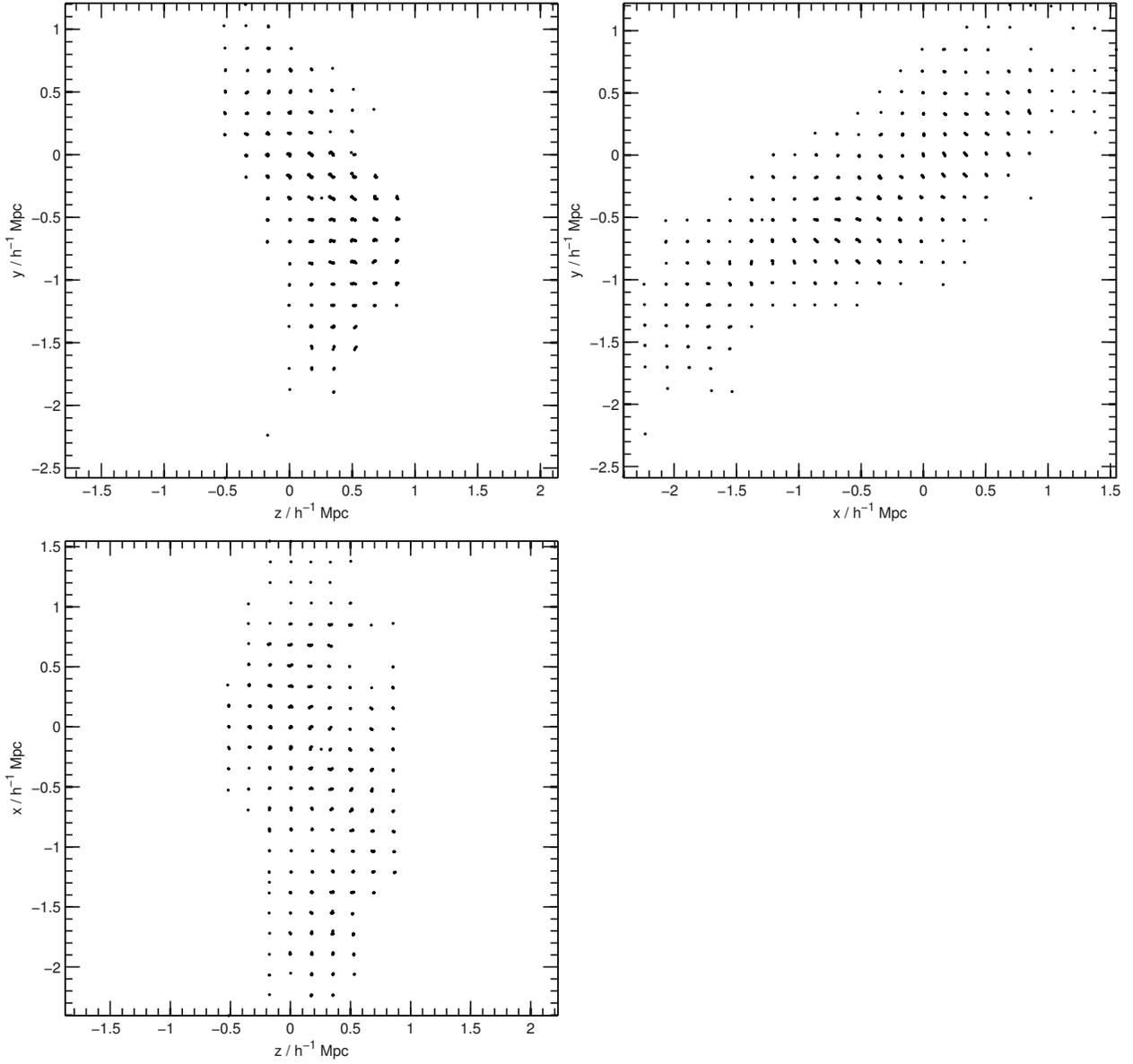}
\caption{Particle distribution of a proto-halo with $S=0.24$ in the 
$x$-$y$ , $y$-$z$ and $x$-$z$ plane in the top-left, top-right and 
bottom-left panel, respectively.}
\label{fig:lowS1}
 \end{center}
\end{figure}

\begin{figure}
  \begin{center}
   \plotone{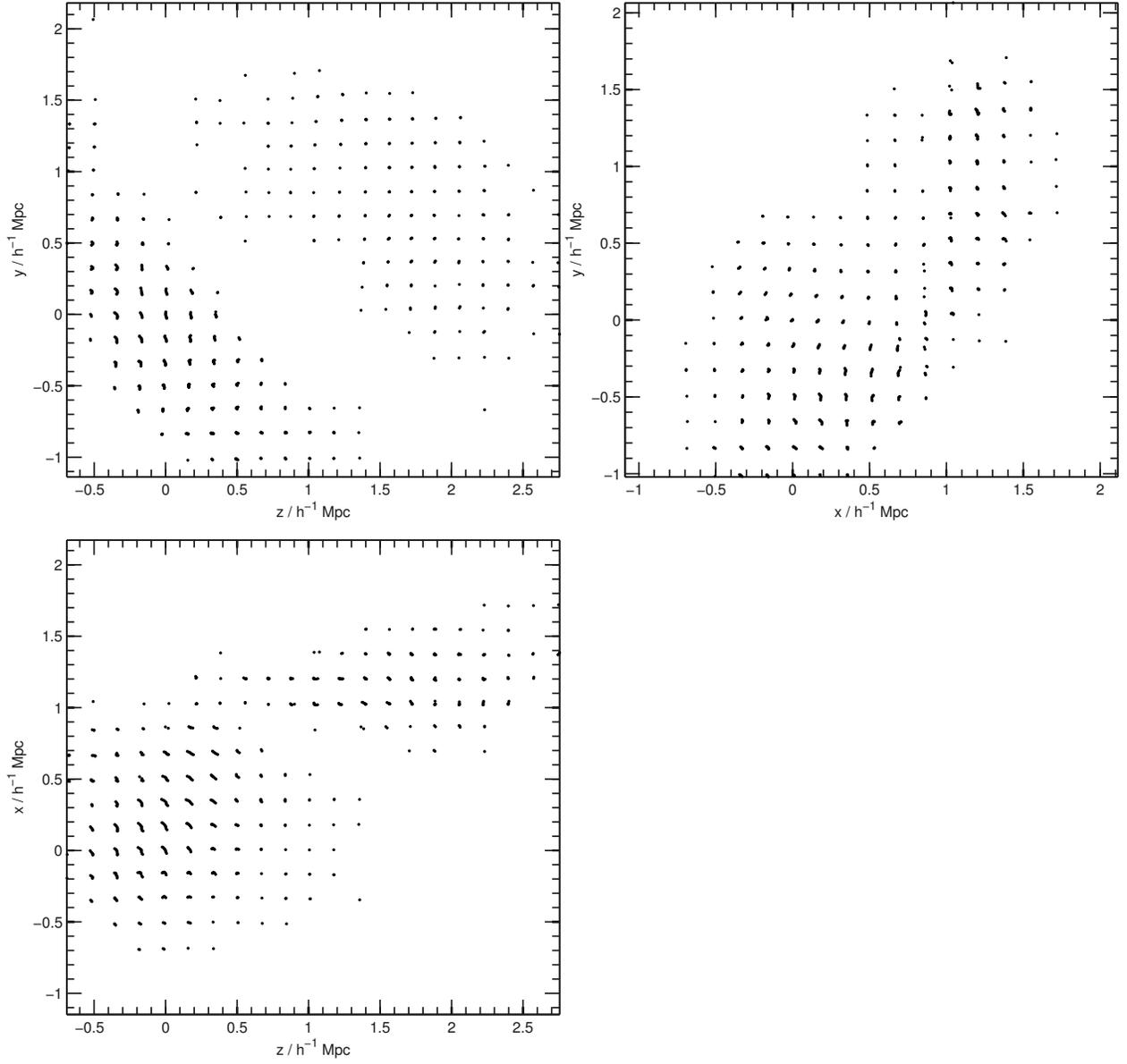}
\caption{Same as Figure \ref{fig:lowS1} but for the case of a disconnected 
proto-halo with $S=0.25$.}
\label{fig:lowS2}
 \end{center}
\end{figure}

Figure \ref{fig:lowS1} shows an example of a connected proto-halo, 
plotting its particle distribution in the $z$-$y$ , $x$-$y$ and $z$-$x$ 
plane in the top-left, top-right and bottom-left panel, respectively.
The sphericity of this example is as low as $S=0.24$. As it can be seen, 
the Lagrangian region is connected, having indeed quite an elongated shape. 
Figure \ref{fig:lowS2} shows an example of the disconnected proto-halo 
regions. Its sphericity is found to be $S=0.25$. Strictly speaking, the 
first order linear tidal torque theory is not valid to explain the 
generation of the angular momentum of such a disconnected proto-halo as 
shown in Figure \ref{fig:lowS2}. In this case it might be the gravitational 
merging of the disconnected patches in the subsequent evolution rather 
than the misalignments between the inertia and tidal tensors in Lagrangian 
space that would contribute to the built-up of the higher angular momentum 
\citep{vit-etal02}. 

Whether or not a proto-halo region is connected, however, the particles of 
a proto-halo region with lower value of $\beta$ (and thus lower values of 
$S$) will end up in a final halo with higher specific angular momentum, 
as revealed in Figure \ref{fig:ang}. Therefore, it is still possible to 
postulate that the peculiar proto-halos with low values of $\beta$ correspond 
to the present LSBGs, no matter what caused the built-up of high angular 
momentum of the LSBGs. 

\section{DISCUSSION AND CONCLUSION}

In the classical Zel'dovich model \citep{zel70}, the inertia tensors of 
bound objects are perfectly correlated with the local tidal tensors in 
Lagrangian space. In practice, the two tensors are found to be indeed 
strongly but not perfectly correlated \citep{lee-pen00,por-etal02b}.
Here, we have determined quantitatively how the strengths of the correlations 
between the two tensors depend on proto-halo's shape and mass, local density 
and filtering scale.  Since the proto-halos form through the tidal flows of 
dark matter particles along the principal axes of the local tidal fields 
from the surrounding matter distribution, it is in fact natural to expect 
strong correlations between the two tensors. Deviations from the perfect 
correlations of the two tensors imply the existence of higher order 
perturbations from the simple tidal flows of CDM particles.

We have also found that for the peculiar proto-halos with low-$S$ the 
correlations between the two tensors tend to be weak (i.e., having low 
value of $\beta$). Since those proto-halos which exhibit lower value of 
$\beta$ end up in halos with higher specific angular momentum, 
it is intriguing to speculate a hypothesis that those 
peculiar proto-halos with lowest values of $S$ would develop into the 
low surface brightness galaxies (LSBGs) at present epoch. 
Since the LSBGs are believed to be dark-matter dominated, their density 
profiles are often directly compared with that of the dark halos 
\citep[i.e., NFW profile][]{nfw96} and the shallow inner-core slope 
of the observed density profiles of LSGBs has been used as a counter-evidence 
for the cold dark matter paradigm \citep[e.g.,][]{moo94}.
For instance, \citet{kuz-etal08} have recently studied the rotation curves 
of $17$ LSBGs obtained from the high-resolution optical velocity fields from 
DensePak spectroscopic observations and shown that the observed LSBGs can 
be matched with the NFW halos only if the LSBGs have $20$ km$s^{-1}$ 
non-circular motions. If the LSBGs originate from the peculiar proto-halo 
sites with exceptionally low value of $\beta$ as in our hypothesis, then 
their proto-halo sites may have had very low sphericity as our results imply.  
Those peculiar proto-halo sites which have extremely low sphericity at the 
initial stages might as well develop non-circular motions. 

It will be interesting to study numerically the density profiles of those 
proto-halos with exceptionally low values of $\beta$ and compare them with the 
standard NFW ones. It will be also interesting to compare the number density 
of those proto-halos with low values of $\beta$ with that of the observed 
LSBGs as a function of mass. We plan to work on these two projects and 
hope to report the results elsewhere in the future.

\acknowledgments

We thank an anonymous referee for a very helpful report.
J.L. acknowledges the financial support from the Korea Science and Engineering 
Foundation (KOSEF) grant funded by the Korean Government 
(MOST, NO. R01-2007-000-10246-0). O.H. acknowledges support from the 
Swiss National Science Foundation. All simulations were performed on the
Gonzales cluster at ETH Zurich, Switzerland.

\clearpage
\clearpage
\clearpage
\clearpage
\clearpage
\clearpage
\clearpage
\clearpage
\clearpage
\begin{deluxetable}{ccc}
\tablewidth{0pt}
\setlength{\tabcolsep}{5mm}
\tablehead{$R_{s}$ & $M_{s}$ & $N_{h}$\\
($h^{-1}$Mpc) &($h^{-1}M_{\odot}$) &}
\tablecaption{Filtering radius ($R_{s}$), the top-hat mass ($M_{s}$) 
enclosed by $R_{s}$, the number of halos ($N_{h}$) with mass in range 
of $[0.9M_{s},1.1M_{s}]$}
\startdata   
$0.5$ & $3.6\times 10^{10}$  & $1049$\\
$1$ & $2.9\times 10^{11}$   & $1472$\\
$2$ & $2.3\times 10^{12}$ & $1835$\\
$5$ & $3.6\times 10^{13}$ & $145$\\
\enddata
\label{tab:list}
\end{deluxetable}

\clearpage
\begin{thebibliography}{}
\bibitem[Bardeen et al.(1986)]{bar-etal86} 
Bardeen, J.~M., Bond, J.~R., Kaiser, N., \& Szalay, A.~S.\ 1986, \apj, 304, 15
\bibitem[Boissier et al.(2003)]{boi-etal03} 
Boissier, S., Monnier Ragaigne, D., Prantzos, N., van Driel, W., Balkowski, C.,
\& O'Neil, K.\ 2003, \mnras, 343, 653 
\bibitem[Bond et al.(1996)]{bon-etal96}
Bond, J.~R., Kofman, L., \& Pogosyan, D.\ 1996, \nat, 380, 603 
\bibitem[Bond \& Myers(1996)]{bon-mye96} 
Bond, J.~R., \& Myers, S.~T.\ 1996, \apjs, 103, 1 
\bibitem[Catelan \& Theuns(1996)]{cat-the96} 
Catelan, P., \& Theuns, T.\ 1996, \mnras, 282, 436 
\bibitem[Davis et al.(1985)]{dav-etal85} 
Davus, M., Efstathious, G., Frenk, C.~S., \& White, S.~D.~M.\ 1985, 
\apj, 292, 371 
\bibitem[Desjacques(2008)]{des08} 
Desjacques, V.\ 2008, \mnras, 388, 638 
\bibitem[Doroshkevich(1970)]{dor70} 
Doroshkevich, A.~G.\ 1970, Astrofizika, 6, 581 
\bibitem[Efstathiou et al.(1985)]{efs-etal85} 
Efstathiou, G., Davis, M., White, S.~D.~M., \& Frenk, C.~S.\ 1985, 
\apjs, 57, 241  
\bibitem[Hahn et al.(2007)]{hah-etal07} 
Hahn, O., Porciani, C., Carollo, C.~M., \& Dekel, A.\ 2007, \mnras, 375, 489 
\bibitem[Hahn et al.(2009)]{hah-etal09} 
Hahn, O., Porciani, C., Dekel, A., \& Carollo, C.~M.\ 2008, arXiv:0803.4211 
\bibitem[Jing(2002)]{jing02} 
Jing, Y.~P.\ 2002, \mnras, 335, L89 
\bibitem[Kaiser(1986)]{kai86} 
Kaiser, N.\ 1986, Inner Space/Outer Space: The Interface between Cosmology 
and Particle Physics, 258 
\bibitem[Katz et al.(1993)]{kat-etal93} 
Katz, N., Quinn, T., \& Gelb, J.~M.\ 1993, \mnras, 265, 689 
\bibitem[Kuzio de Naray et al.(2008)]{kuz-etal08} 
Kuzio de Naray, R., McGaugh, S.~S., \& de Blok, W.~J.~G.\ 2008, \apj, 676, 920 
\bibitem[Lee \& Pen(2000)]{lee-pen00} 
Lee, J., \& Pen, U.-L.\ 2000, \apjl, 532, L5 
\bibitem[Maddox et al.(1990)]{mad-etal90}
Maddox, S.~J., Efstathiou, Sutherland, G. W.~J. \& Loveday, J.\ 1990,   
\mnras, 242, 43P
\bibitem[Monnier Ragaigne et al.(2003)]{mon-etal03}
Monnier Ragaigne, D., van Driel, W., Schneider, S.~E., Jarrett, T.~H., 
\& Balkowski, C.\ 2003, \aap, 405, 99
\bibitem[Moore(1994)]{moo94} 
Moore, B.\ 1994, \nat, 370, 629 
\bibitem[Navarro et al.(1996)]{nfw96} 
Navarro, J.~F., Frenk, C.~S., \& White, S.~D.~M.\ 1996, \apj, 462, 563 
\bibitem[Porciani et al.(2002a)]{por-etal02a} 
Porciani, C., Dekel, A., \& Hoffman, Y.\ 2002, \mnras, 332, 325 
\bibitem[Porciani et al.(2002b)]{por-etal02b} 
Porciani, C., Dekel, A., \& Hoffman, Y.\ 2002, \mnras, 332, 339 
\bibitem[Roberts et al.(2007)]{rob-etal07}
Roberts, S., Davies, J., Sabatini, S., Auld, R., \& Smith, R.\ 2007, 
\mnras, 379, 1053 
\bibitem[Robertson et al.(2009)]{rob-etal09} 
Robertson, B.~E., Kravtsov, A.~V., Tinker, J., \& Zentner, A.~R.\ 2009, 
\apj, 696, 636 
\bibitem[Sheth et al.(2001)]{she-etal01} 
Sheth, R.~K., Mo, H.~J., \& Tormen, G.\ 2001, \mnras, 323, 1 
\bibitem[Vitvitska et. al.(2002)]{vit-etal02} 
Vitvitska, M., Klypin, A. A., Kravtsov, A. V., Wechsler, R. H., 
Primack, J. R., \& Bullock, J. S. 2002, \apj, 581, 799
\bibitem[White(1984)]{whi84} 
White, S.~D.~M.\ 1984, \apj, 286, 38
\bibitem[Zel'dovich(1970)]{zel70}
Zel'dovich, Ya. B.\ 1970, \aap, 5, 84
\end{thebibliography}
\end{document}